\documentclass[preprint]{vgtc}               

\graphicspath{{figures/}{pictures/}{images/}{./}} 

\usepackage{times}                     

\usepackage{tabu}                      
\usepackage{booktabs}                  
\usepackage{lipsum}                    
\usepackage{mwe}                       

\usepackage{mathptmx}                  
\usepackage{amsmath}
\usepackage{amssymb}
\usepackage{comment}
\usepackage{multirow}
\usepackage{graphicx}

\usepackage{subcaption}
\onlineid{1861}

\vgtccategory{Research}

\vgtcinsertpkg

\preprinttext{This is a preprint version of the manuscript.}

\title{Beyond Monoscopic Viewing: \\ A Study on 3D Gaussian Splatting Quality in VR}

\author{Shreyas Shivakumara\thanks{e-mail: shreyas.shivakumara@liu.se}%
\and Gabriel Eilertsen\thanks{e-mail: gabriel.eilertsen@liu.se}%
\and Karljohan Lundin Palmerius\thanks{e-mail: karljohan.lundin.palmerius@liu.se}}

\affiliation{\scriptsize
Department of Science and Technology, Linköping University, Norrköping, Sweden}

\teaser{
  \centering
  \includegraphics[width=\textwidth]{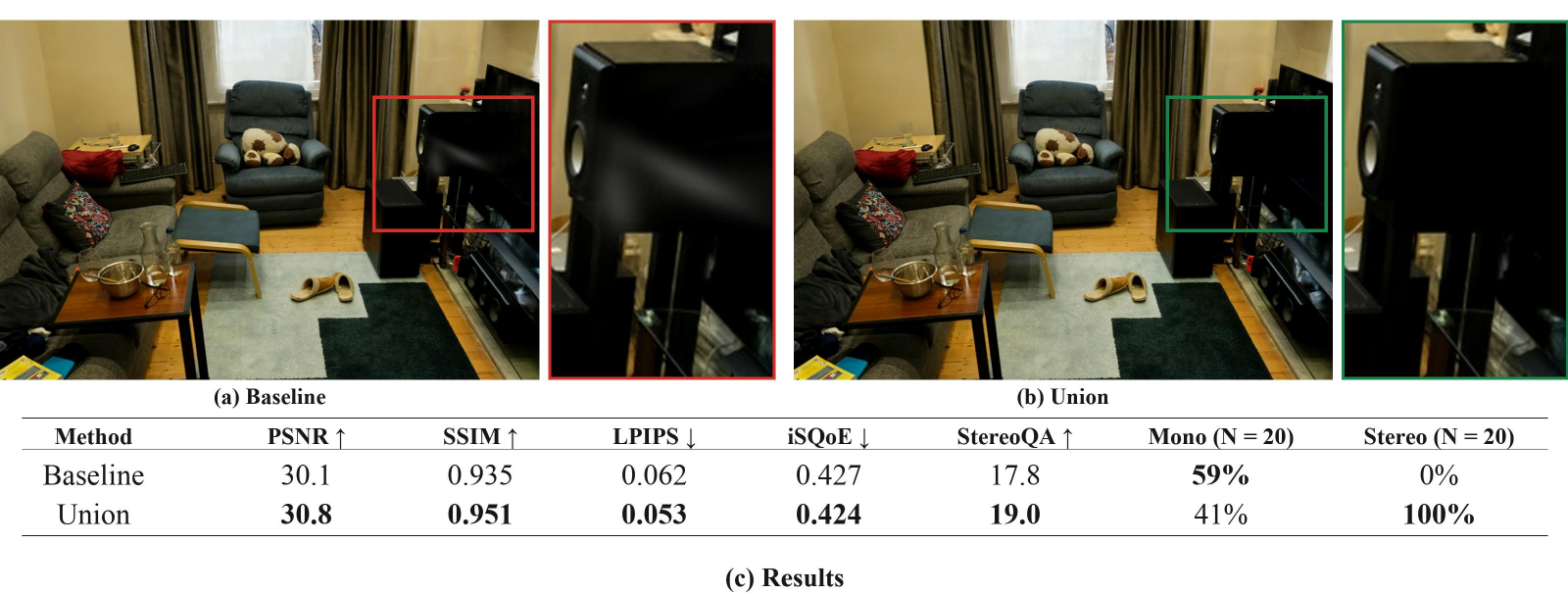}
  \caption{ 
  Comparison of baseline (SfM-only) and Union initialization on the \textit{room} scene. (a)~Baseline rendering with a zoomed region (red box) showing a floater artifact.  (b)~Union initialized rendering with the corresponding region (green box) is showing reconstruction without the floater. (c)~Quantitative metrics and human preference between the two reconstructions. Standard image metrics (PSNR$\uparrow$, SSIM$\uparrow$, LPIPS$\downarrow$) and stereo-aware metrics (iSQoE$\downarrow$, StereoQA$\uparrow$) show marginal differences between the two initializations. Monoscopic preference is near chance ($41\%$ for the union, $N = 20$), while stereoscopic VR viewing shows a pronounced difference ($100\%$ for the union, $N = 20$). Standard 2D metrics and monoscopic viewing miss stereo-specific artifacts that become perceptually salient under stereo viewing. Arrows indicate whether higher or lower values are better in metrics.
  }
  \label{fig:firstimage}
}

\abstract{
    Stereoscopy is fundamental to virtual reality (VR), providing depth perception through binocular viewing. 
    Recent advances in 3D Gaussian Splatting (3DGS) enable high-quality novel view synthesis, making it well suited to immersive VR. 
    We render 3DGS reconstructions stereoscopically and evaluate them in a head-mounted display, replicating how they would actually be viewed in VR.
    Real-world capture provides only a limited number of views, and under this constraint 3DGS reconstruction often produces localized floaters and misplaced structures. 
    Standard metrics miss these localized artifacts, which become salient under stereoscopic viewing, where geometry is placed at the wrong depth.

    We investigate whether standard image-quality evaluation reflects the perceptual quality of 3DGS reconstructions under reduced capture. 
    We compare SfM-only baseline with a union initialization that combines SfM with a dense VGGT network. 
    All other training components are held fixed, isolating the effect of initialization coverage.
    We conduct a user study comparing preferences under monoscopic and stereoscopic HMD viewing, and test whether image-quality metrics predict the observed preferences.
    Monoscopically, preference for the more consistent reconstruction is weak, reaching 58.4\% overall. Stereoscopically, the same preference rises to 78.2\% and is consistent across all participants, while image-quality metrics (PSNR, SSIM and LPIPS) and stereo-aware metrics (iSQoe and StereoQA) show only modest differences and fail to penalize them. Our results indicate that monoscopic evaluation and standard image-quality metrics substantially underestimate perceptual artifacts observed in 3DGS reconstructions for VR, making stereoscopic assessment essential for 3DGS quality evaluation in VR.

    } 

\keywords{Gaussian Splatting, Stereoscopy, Evaluation}

\begin{document}


\firstsection{Introduction}
\maketitle

3D Gaussian Splatting (3DGS) has revolutionized novel-view synthesis, setting a new standard for the visual quality of rendered images~\cite{kerbl20233d}. It produces high-quality, photorealistic results with short training times and renders them in real time, opening opportunities for many applications. One such application is virtual reality (VR), where a captured scene can be explored immersively and each eye is shown its own view, so that depth arises from binocular disparity. In this work, we investigate 3DGS reconstruction for consumer head-mounted displays (HMD).

3DGS assumes a dense capture with substantial overlap and broad coverage of the scene from all angles. Real-world captures rarely satisfy these conditions. Vanilla 3DGS relies on Structure-from-Motion (SfM) to seed the initial points that optimization then refines~\cite{schonberger2016structure}. As the number of views decreases, this seed cloud becomes sparse and uneven, producing floaters and loss of texture~\cite{barron2022mip, Nerfbusters2023}. Such artifacts are problematic in VR because incorrect geometry may cause incorrect projections or disparities across the two eyes.

Nevertheless, 3DGS reconstruction quality is commonly evaluated using PSNR, SSIM, and LPIPS on held-out monoscopic views. These metrics aggregate image differences over pixels and do not explicitly model binocular consistency or perceived depth. Spatially localized artifacts in 3DGS reconstructions occupy a small fraction of the image and receive a modest penalty, while being particularly salient in stereoscopic viewing. This is also observed in other work where there is divergence between conventional image metrics and subjective assessment of neural rendering quality~\cite{liang2024perceptual, martin2024nerf}.

We address this question through a controlled comparison of two 3DGS initialization strategies under reduced capture replicating real-world settings. The baseline initializes 3DGS using SfM points alone, whereas the proposed union initialization combines the SfM points with a dense feed-forward geometric prior. All subsequent components of the pipeline including camera poses, architecture, optimizer, hyperparameters, and training schedule are held identical. The resulting reconstruction pairs therefore isolate the effect of the initialization strategy while avoiding confounding changes elsewhere in the pipeline.

In this paper, we study and evaluate 3DGS stereoscopically in a consumer VR headset replicating real-world settings. We evaluate the reconstruction pairs quantitatively and with a user study. We measure held-out image quality with standard image metrics (PSNR, SSIM, and LPIPS). Next we render stereo pairs and evaluate stereo quality using stereo-aware metrics (iSQoE and StereoQA), which are trained to detect stereoscopic artifacts. Third, we conduct a two-alternative forced-choice user study in which separate participant groups compare the reconstructions under monoscopic and stereoscopic viewing conditions. This design allows us to determine whether metric improvements correspond to perceptual preferences and whether those preferences change when binocular depth cues are introduced. We show that artifacts observed in reduced-capture 3DGS remain inconspicuous in monoscopic images and are weakly reflected in conventional metrics but are salient under stereoscopic viewing. We therefore recommend that if 3DGS reconstructions are intended for VR, their quality should be evaluated perceptually and in HMDs.

\begin{figure*}[t]
 \centering
 \includegraphics[width=\textwidth]{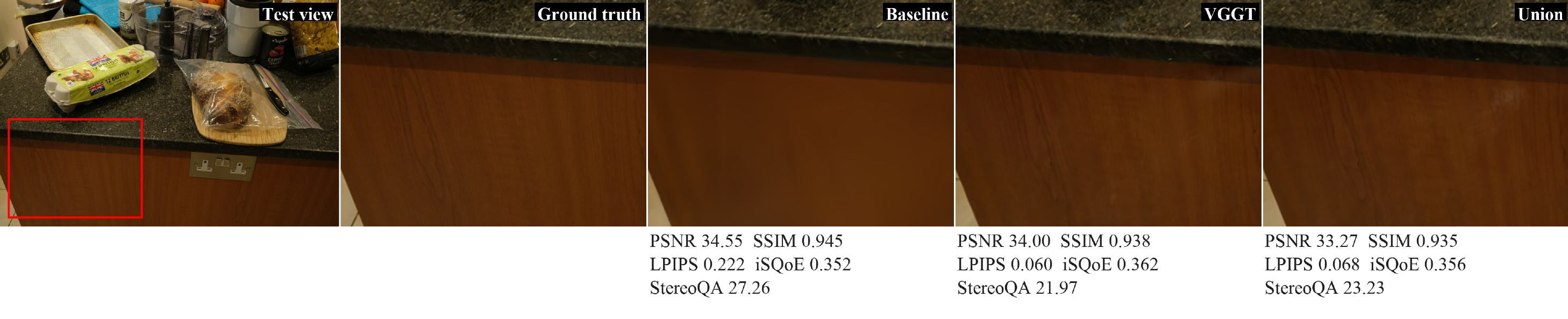}\\[4pt]
 \includegraphics[width=\textwidth]{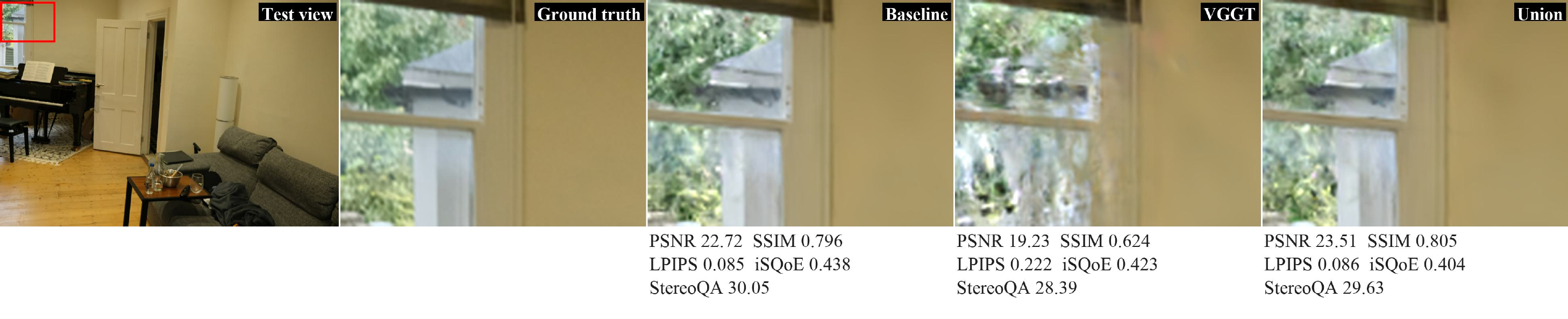}
 \caption{The red box marks the region of interest and the four right panels show zoomed crops for the ground truth, baseline (SfM-only initialization), VGGT, and Union initialization. In the top scene, the baseline method struggles to render near-field wood-grain texture on the cabinet. The texture is smoothed out, while VGGT and union restore it closer to the ground truth. In the bottom scene, VGGT shows a ghosting artifact in the far-field window and foliage, which the baseline and union resolve with sharper edges that look closer to the ground truth. Despite this perceptually large difference, standard 2D metrics and stereo-specific metrics, iSQOE shows a marginal difference but the StereoQA shows good improvement in this scene. This suggests that metric evaluation is largely insensitive to localized and loss of texture artifacts. When viewed in an HMD, however, such artifacts are amplified and are not preferred by users.}
 \label{problem_definition_image}
\end{figure*}

\section{Related Work} 

We review related work on 3DGS, its use in virtual reality, evaluation of novel-view synthesis, and initialization in gaussian-splatting. 

\textbf{3D Gaussian Splatting} Implicit 3D scene representations using volumetric fields became popular since the introduction of neural radiance fields (NeRF)~\cite{mildenhall2021nerf}. Building on this success, 3DGS~\cite{kerbl20233d} instead represents a scene explicitly as a set of anisotropic 3D Gaussians, enabling real-time novel-view synthesis. By combining these Gaussians with a fast differentiable rasterizer, it achieves real-time rendering at high visual fidelity. 3D Gaussian Splatting has been adopted across dynamic and 4D scene modeling~\cite{wu20244d}, human avatar reconstruction~\cite{qian2024gaussianavatars, li2024animatable} and 3D generation, and editing~\cite{tang2024dreamgaussian}. In all of these applications, reconstruction quality is almost always evaluated by the image metrics computed on monoscopic viewing.

\textbf{3DGS for Virtual Reality} There is growing interest in adapting radiance-field representations for head-mounted display (HMD) viewing. One direction is efficient, foveated rendering that exploits gaze to concentrate compute where the viewer looks. VR-Splatting~\cite{franke2025vr} combines foveated rendering with a point-based representation to improve performance in the headset, and Fov-GS~\cite{fan2025fov} renders dynamic scenes in real time with a foveated 3D Gaussian representation to improve perceptual quality in the foveal region. Another direction addresses the artifacts observed in HMDs. VRSplat~\cite{tu2025vrsplat} combines several rasterization improvements to suppress popping and distortion during head motion. Beyond rendering, interactive systems embed Gaussian splatting in VR for physically based manipulation~\cite{jiang2024vr}, and further work extends it to domains such as medical visualization~\cite{xia2026adaendogs, Kleinbeck_2026} and language-driven interaction in immersive environments~\cite{mao2026live}. While these works have demonstrated that Gaussian Splatting can be used for VR, they focus on rendering-stage or interaction improvements and typically assume high-quality input reconstructions. We ask whether vanilla 3DGS can deliver a convincing stereoscopic experience in an HMD under reduced view coverage.

\textbf{Evaluation of Novel-View Synthesis} Evaluating a 3DGS reconstruction is essential for assessing the model's quality. The standard evaluation compares rendered held-out views against ground-truth images using image metrics. PSNR, SSIM~\cite{wang2004image}, and LPIPS~\cite{lpips2018} are the standard metrics for novel-view synthesis and are adopted across the field~\cite{kerbl20233d, Yu2023MipSplatting, lu2024scaffold}. These metrics, however, are computed monocularly and as full-frame averages. 
Prior subjective studies have shown this divergence between such metrics and human judgment in novel-view synthesis conducted on monoscopic displays~\cite{martin2025gs, liang2024perceptual, shivakumara2025exploring, martin2024nerf}. Since our concern is stereoscopic quality, we must account for binocular disparity conflicts and localized errors that are typically missed under monoscopic evaluation. The closest domain-appropriate measure is iSQoE~\cite{tamir2025makes}, a learned stereoscopic quality metric trained on reconstruction-induced artifacts in addition to natural-image distortions, which we therefore also report. Yet stereoscopic quality is ultimately a perceptual property of the rendered image pair, and no existing metric is established to capture the localized, floater artifacts that arise under reduced view coverage. We therefore conduct a two-alternative forced-choice (2AFC) study both on a monoscopic display and in an HMD, following the protocol for stereoscopic Gaussian splatting in VRSplat~\cite{tu2025vrsplat}, directly measuring the human preference that image- and stereo-quality metrics aim to predict.

\textbf{Initialization for Novel-View Synthesis} 
3DGS typically requires dense capture of the scene. Synthesizing novel views from fewer images has therefore attracted significant interest in the community. Three broad strategies have emerged. The first injects depth priors as additional supervision during training~\cite{zhu2024fsgs, li2024dngaussian, xiong2025sparsegs}. The second replaces traditional SfM with differentiable or learned pipelines~\cite{wang2024vggsfm, duisterhof2025mast3r, lee2025dense, li2025fastmap}. The third initializes from feed-forward geometry models such as DUSt3R~\cite{wang2024dust3r}, MASt3R~\cite{leroy2024grounding}, and VGGT~\cite{wang2025vggt}, which regress dense 3D structure from few or unposed images. Subsequent work uses these predictions to initialize or constrain 3DGS~\cite{jiang2025anysplat, xiang2026vggs}.
Most closely related to our initialization approach, SPARS3R~\cite{tang2025spars3r} fuses feed-forward dense reconstruction with SfM points through correspondence-based warping, while InstantSplat~\cite{instantsplat} reconstructs scenes from 2D images using MASt3R priors. Closest to our stereoscopic framing, \cite{han2024binocular} exploit binocular consistency for sparse-view synthesis. However, all of these approaches modify the vanilla 3DGS pipeline, replacing SfM, injecting dense geometric or binocular priors directly into optimization, or both, and none are evaluated in VR. We instead use a dense prior solely to seed the initial point cloud and produce the controlled stimuli for the evaluation, leaving the vanilla 3DGS architecture unchanged. 

\section{Problem Formulation} 
\label{problem_statement}

\begin{figure*}[t]
 \centering
 \includegraphics[width=\textwidth]{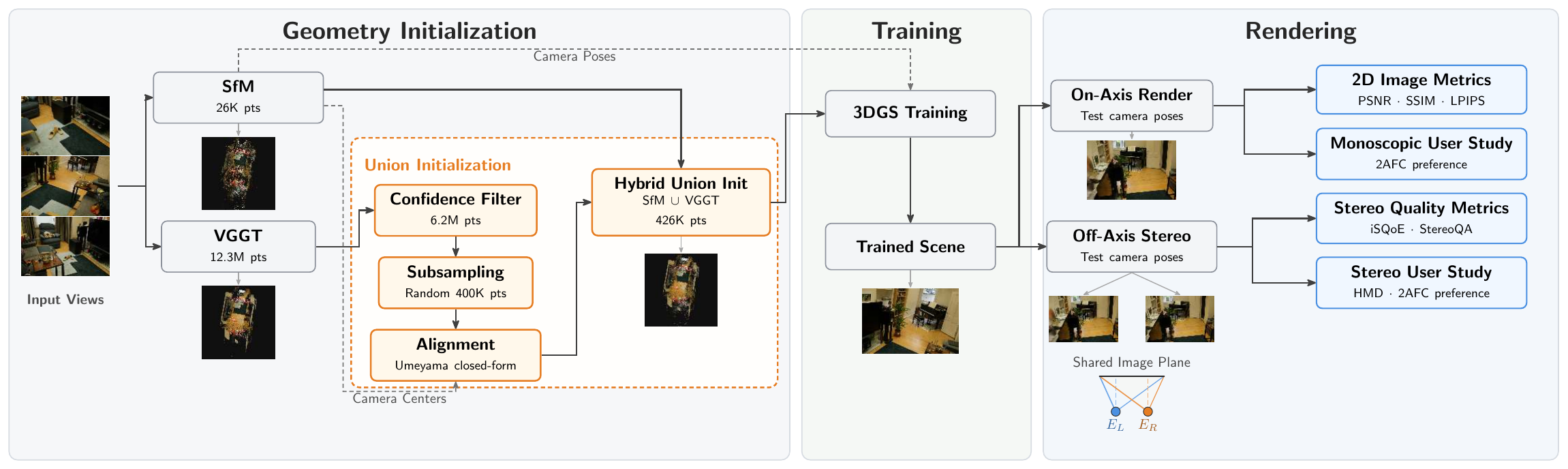}
 \caption{\textbf{Overview of Union initialization pipeline.} (a) Geometry initialization: SfM points are fused with a filtered, subsampled, and aligned VGGT dense prior to produce a hybrid point cloud. (b) Training: The hybrid point cloud initializes standard 3DGS with fixed SfM camera poses. (c) Evaluation: The trained scene is rendered for image-quality metrics and a monoscopic user study (on-axis), and for a stereoscopic user study in a HMD (off-axis stereo pairs).}
 \label{fig:architecture}
\end{figure*}

3DGS assumes dense captures of the scene, acquiring a few hundred images that cover all parts of the scene \cite{kerbl20233d, lu2024scaffold}. 
However, this is rarely what practical acquisition looks like. 
A user generally captures a handful of images, which produces far fewer views of the scene.  
We study 3DGS~\cite{kerbl20233d} for novel-view synthesis under reduced scene coverage, and evaluate the result stereoscopically in a consumer-grade HMD. 
In an HMD the scene is rendered as a stereo pair, one image per eye, so that binocular disparity conveys depth~\cite{wheatstone1838contributions, banks2012stereoscopy, howard2012perceiving}.
To replicate the real-world setting, we model this by uniform stride sampling where we train the 3DGS model on every 4th, 8th, and 16th frame of a standard capture sequence.
This spans the whole scene but from far fewer viewpoints. 
 
3DGS relies on Structure-from-Motion (SfM)~\cite{schonberger2016structure} to produce the initial point cloud. 
These seed points are where 3DGS places its initial Gaussians. 
Under reduced coverage, this cloud becomes sparse and unevenly distributed. 
Regions with little coverage receive few or no seed points, and densification during training cannot recover geometry it was never seeded with.
This results in artifacts such as floaters at the wrong depth or loss of texture~\cite{Nerfbusters2023, barron2022mip}, as shown in Figure~\ref{problem_definition_image}. The texture-less regions in the SfM example give the visual system nothing to match, so the depth there cannot be recovered and the artifact is clearly seen in the headset~\cite{julesz1971foundations}. 
A natural hypothesis is therefore to replace the SfM cloud with a dense feed-forward prediction that assigns a 3D point to every pixel of every input view. 
We therefore consider a feed-forward model such as VGGT~\cite{wang2025vggt}, which produces dense seed points that cover the entire scene, including the regions SfM leaves empty. However, these predictions are accurate in the near-field but struggle in the far-field, as seen in Figure~\ref{problem_definition_image}, where the window appears hazy and heavily blurred. 
The far-field errors in the VGGT initialization place geometry at the wrong depth, which causes discomfort and breaks the sense of depth~\cite{kooi2004visual, zhu2025perceptual}. 

Neither initialization source alone is sufficient. We therefore construct pairs of reconstruction to examine the union of the two initialization strategy affects stereoscopic quality, keeping all other factors identical to create controlled stimuli.

Reduced coverage produces artifacts that standard 2D metrics barely register and that monoscopic viewing can miss, yet they are pronounced in HMDs. This metric insensitivity motivates our perceptual study (Section~\ref{sec:user_study}), in which observers judge identical content under both stereoscopic HMD and monoscopic conditions in a 2AFC protocol. Isolating the cause of these artifacts requires controlled stimuli. We therefore construct them in Section~\ref{sec:controlled_stimulus} by confining all changes to initialization. The training parameters, architecture, and optimization remain identical to baseline 3DGS, so any differences between reconstructions are attributed to seed geometry alone.

\begin{figure}[t]
 \centering
  \includegraphics[width=\columnwidth]{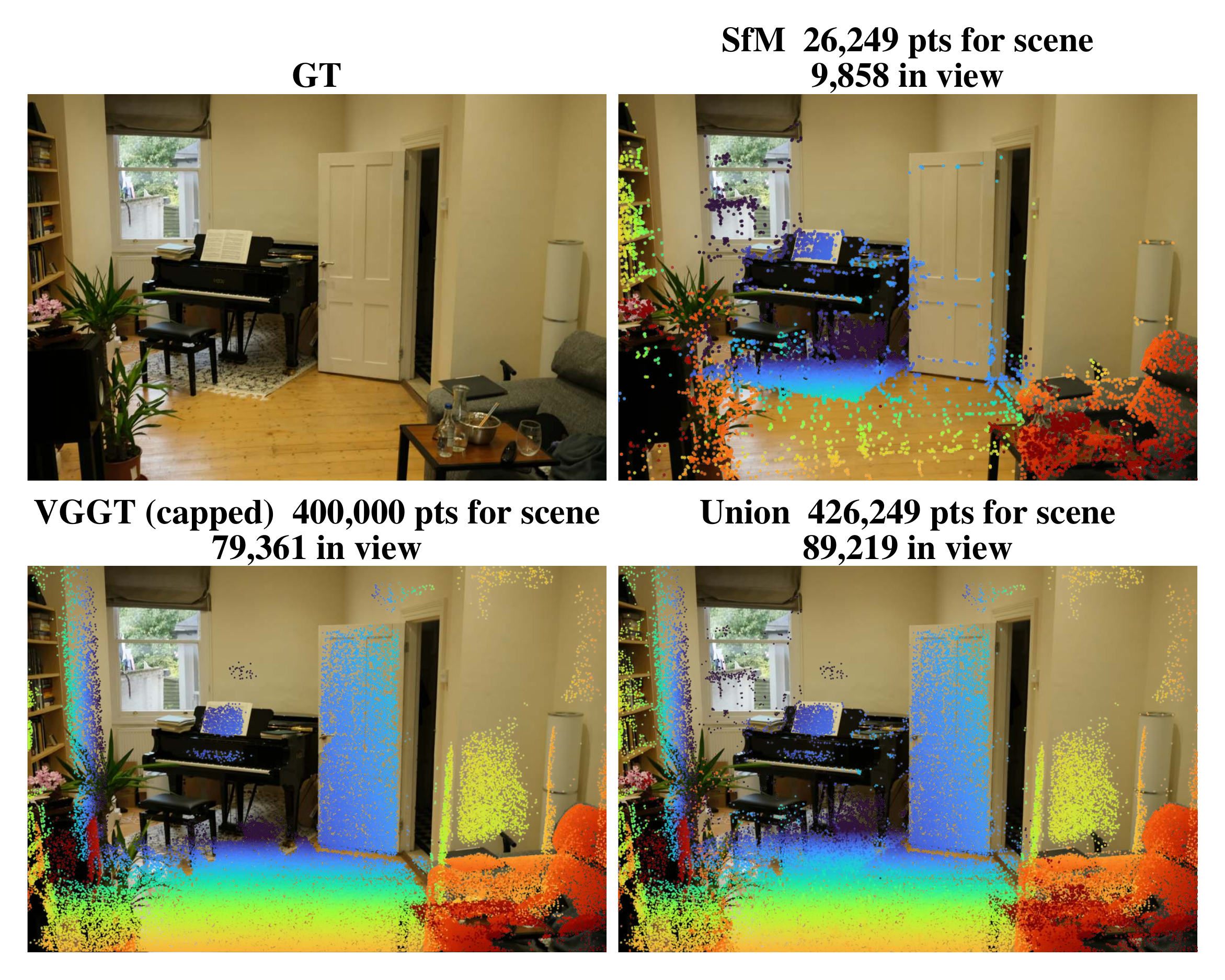}
 \caption{Illustrating an example in this view where VGGT alone misses certain far-field points (window seed points) that SfM captures, yet SfM 9,858 points are insufficient for Gaussian splatting. Union combines both for denser seed point initialization (89,219 points).}
  \label{fig:init_coverage}
\end{figure}

\begin{table*}[t]
\centering
\caption{\textbf{Quantitative comparison of reconstruction quality}. Standard image metrics (PSNR, SSIM, and LPIPS) are evaluated on held-out on-axis views. Stereo-aware metrics (iSQoE and StereoQA) evaluated on off-axis stereo pairs. The mean rows indicate the average across the eight scenes in each data capture settings. Arrows indicate the preferred direction of the magnitude. \textbf{Bold} indicates the best score per metric. The benefit of the union initialization is seen clearly as the capture density decreases except the StereoQA metric.}
\label{tab:quantitative_results}
\scriptsize
\setlength{\tabcolsep}{2.5pt}
\begin{tabular}{l ccc ccc ccc ccc ccc}
\toprule
\multirow{2}{*}{Scene} & \multicolumn{3}{c}{PSNR $\uparrow$} & \multicolumn{3}{c}{SSIM $\uparrow$} & \multicolumn{3}{c}{LPIPS $\downarrow$} & \multicolumn{3}{c}{iSQoE $\downarrow$} & \multicolumn{3}{c}{StereoQA $\uparrow$} \\
\cmidrule(lr){2-4}\cmidrule(lr){5-7}\cmidrule(lr){8-10}\cmidrule(lr){11-13}\cmidrule(lr){14-16}
 & SfM & VGGT & Union & SfM & VGGT & Union & SfM & VGGT & Union & SfM & VGGT & Union & SfM & VGGT & Union \\
\midrule
\multicolumn{16}{l}{\textbf{1/4 Views}} \\
\midrule
Bicycle  & 20.53 & 20.42 & \textbf{21.54} & 0.457 & 0.543 & \textbf{0.571} & 0.406 & 0.253 & \textbf{0.223} & 0.567 & 0.512 & \textbf{0.493} & \textbf{13.91} & 7.16 & 7.09 \\
Bonsai   & \textbf{30.71} & 28.89 & 30.37 & \textbf{0.939} & 0.899 & 0.935 & 0.064 & 0.063 & \textbf{0.052} & \textbf{0.442} & 0.460 & 0.445 & \textbf{25.15} & 23.63 & 23.82 \\
Counter  & 25.76 & 25.73 & \textbf{26.10} & 0.851 & 0.847 & \textbf{0.858} & 0.114 & 0.100 & \textbf{0.096} & 0.505 & 0.503 & \textbf{0.494} & \textbf{21.00} & 19.30 & 19.62 \\
Garden   & 24.74 & 22.67 & \textbf{24.78} & 0.793 & 0.733 & \textbf{0.798} & 0.115 & 0.132 & \textbf{0.100} & 0.428 & 0.437 & \textbf{0.417} & \textbf{7.28} & 5.96 & 6.01 \\
Kitchen  & \textbf{27.52} & 23.22 & 27.25 & \textbf{0.917} & 0.853 & 0.911 & \textbf{0.070} & 0.125 & 0.074 & \textbf{0.477} & 0.536 & 0.481 & \textbf{17.88} & 17.21 & 17.68 \\
Playroom & 23.89 & 25.28 & \textbf{25.43} & 0.828 & 0.849 & \textbf{0.850} & 0.259 & 0.214 & \textbf{0.211} & 0.802 & \textbf{0.767} & 0.767 & 34.17 & \textbf{34.73} & 34.48 \\
Room     & 29.82 & 29.64 & \textbf{30.52} & 0.924 & 0.919 & \textbf{0.925} & 0.085 & 0.078 & \textbf{0.074} & 0.591 & \textbf{0.577} & 0.581 & \textbf{21.47} & 21.09 & 20.80 \\
Stump    & 17.75 & 21.51 & \textbf{21.61} & 0.350 & 0.488 & \textbf{0.491} & 0.462 & 0.291 & \textbf{0.284} & 0.818 & 0.757 & \textbf{0.752} & \textbf{15.50} & 9.87 & 9.68 \\
\midrule
Mean & 25.09 & 24.67 & \textbf{25.95} & 0.757 & 0.766 & \textbf{0.792} & 0.197 & 0.157 & \textbf{0.139} & 0.579 & 0.569 & \textbf{0.554} & \textbf{19.55} & 17.37 & 17.40 \\
\midrule
\multicolumn{16}{l}{\textbf{1/8 Views}} \\
\midrule
Bicycle  & 15.27 & 19.00 & \textbf{19.06} & 0.212 & 0.419 & \textbf{0.419} & 0.503 & 0.321 & \textbf{0.317} & 0.683 & \textbf{0.564} & 0.577 & \textbf{7.81} & 6.59 & 6.76 \\
Bonsai   & 22.50 & \textbf{24.48} & 23.92 & 0.735 & \textbf{0.768} & 0.759 & 0.195 & \textbf{0.132} & 0.139 & \textbf{0.542} & 0.588 & 0.582 & \textbf{23.27} & 22.29 & 22.27 \\
Counter  & 22.11 & \textbf{22.36} & 22.31 & 0.737 & \textbf{0.751} & 0.749 & 0.179 & \textbf{0.152} & 0.159 & 0.600 & 0.593 & \textbf{0.589} & \textbf{19.97} & 18.76 & 18.90 \\
Garden   & 20.54 & 20.16 & \textbf{21.29} & 0.608 & 0.595 & \textbf{0.644} & 0.214 & 0.198 & \textbf{0.173} & 0.485 & 0.479 & \textbf{0.452} & \textbf{7.49} & 6.01 & 6.32 \\
Kitchen  & \textbf{23.08} & 20.22 & 23.66 & \textbf{0.847} & 0.755 & 0.834 & 0.128 & 0.225 & \textbf{0.125} & 0.529 & 0.606 & \textbf{0.528} & 18.79 & \textbf{19.70} & 17.88 \\
Playroom & 15.76 & \textbf{18.88} & 18.88 & 0.624 & \textbf{0.726} & 0.726 & 0.517 & \textbf{0.351} & 0.351 & 0.871 & \textbf{0.848} & 0.850 & 31.58 & 32.91 & \textbf{33.30} \\
Room     & 22.82 & 26.57 & \textbf{26.71} & 0.783 & \textbf{0.856} & 0.854 & 0.210 & \textbf{0.119} & 0.121 & 0.714 & \textbf{0.609} & 0.619 & \textbf{22.57} & 20.55 & 20.38 \\
Stump    & 15.98 & \textbf{19.39} & 19.31 & 0.186 & 0.357 & \textbf{0.359} & 0.510 & 0.396 & \textbf{0.391} & 0.854 & 0.829 & \textbf{0.818} & \textbf{13.36} & 10.62 & 10.61 \\
\midrule
Mean & 19.88 & 21.38 & \textbf{21.89} & 0.592 & 0.653 & \textbf{0.668} & 0.307 & 0.237 & \textbf{0.222} & 0.660 & 0.640 & \textbf{0.627} & \textbf{18.11} & 17.18 & 17.05 \\
\midrule
\multicolumn{16}{l}{\textbf{1/16 Views}} \\
\midrule
Bicycle  & 13.47 & 16.42 & \textbf{16.56} & 0.132 & \textbf{0.292} & 0.292 & 0.608 & 0.428 & \textbf{0.423} & 0.805 & \textbf{0.689} & 0.692 & \textbf{7.93} & 7.13 & 7.23 \\
Bonsai   & 15.88 & \textbf{19.25} & 19.13 & 0.411 & \textbf{0.567} & 0.564 & 0.486 & \textbf{0.283} & 0.290 & 0.814 & 0.746 & \textbf{0.743} & 19.75 & 20.99 & \textbf{21.16} \\
Counter  & 16.62 & 18.23 & \textbf{18.51} & 0.466 & 0.576 & \textbf{0.587} & 0.401 & 0.296 & \textbf{0.275} & 0.792 & 0.752 & \textbf{0.742} & 18.90 & \textbf{19.27} & 18.45 \\
Garden   & 16.56 & 18.39 & \textbf{18.48} & 0.318 & 0.473 & \textbf{0.473} & 0.422 & 0.278 & \textbf{0.277} & 0.643 & 0.544 & \textbf{0.538} & \textbf{8.55} & 6.83 & 6.93 \\
Kitchen  & \textbf{19.63} & 18.99 & 19.34 & 0.694 & 0.681 & \textbf{0.701} & 0.233 & 0.266 & \textbf{0.227} & 0.687 & \textbf{0.671} & 0.675 & 17.75 & \textbf{19.71} & 17.92 \\
Playroom & 11.30 & 14.46 & \textbf{14.51} & 0.456 & \textbf{0.603} & 0.597 & 0.707 & 0.518 & \textbf{0.505} & \textbf{0.862} & 0.878 & 0.877 & 28.76 & 31.79 & \textbf{31.85} \\
Room     & 16.50 & 22.53 & \textbf{22.68} & 0.539 & 0.753 & \textbf{0.754} & 0.427 & 0.208 & \textbf{0.208} & 0.821 & \textbf{0.692} & 0.693 & \textbf{21.17} & 19.75 & 19.96 \\
Stump    & 11.60 & 15.57 & \textbf{15.70} & 0.017 & 0.169 & \textbf{0.178} & 0.973 & \textbf{0.566} & 0.566 & \textbf{0.756} & 0.844 & 0.848 & \textbf{33.82} & 16.53 & 16.81 \\
\midrule
Mean & 15.20 & 17.98 & \textbf{18.11} & 0.378 & 0.514 & \textbf{0.518} & 0.532 & 0.355 & \textbf{0.346} & 0.773 & 0.727 & \textbf{0.726} & \textbf{19.58} & 17.75 & 17.54 \\
\bottomrule
\end{tabular}
\end{table*}

\section{Controlled Stimulus Generation}
\label{sec:controlled_stimulus}

As described in Section~\ref{problem_statement}, SfM produces accurate but sparse seed points, while VGGT produces dense seed points that degrade in the far field. We aggregate both initial point clouds and feed them to training. All subsequent training, parameters, and optimizations are kept identical to vanilla 3DGS.

\subsection{Preliminaries: 3D Gaussian Splatting}
\label{sec:prelim}

3DGS represents a scene as a set of anisotropic 3D Gaussians and renders novel views in real time via differentiable rasterization. Given a set of input images, camera poses are estimated and a sparse point cloud is triangulated using SfM. Each Gaussian is initialized in this cloud and its position, covariance, opacity, and view-dependent color are optimized against the ground-truth images, with adaptive densification and pruning. Densification operates on Gaussians whose accumulated gradient magnitude exceeds a threshold. They either clone large Gaussians or split them into smaller ones when their accumulated gradient magnitude exceeds a threshold. Meanwhile, Gaussians with near-zero opacity are also pruned during the optimization. Densification only refines or subdivides existing Gaussians, so it cannot introduce structure in regions where the initial SfM cloud is empty. The initial point cloud therefore determines which scene regions are reconstructed during the training process.

\subsection{Hybrid Initialization}

\textbf{Dense prediction.}
We use VGGT~\cite{wang2025vggt}, a large feed-forward transformer that maps the $N$ input images to per-view 3D annotations,
\begin{equation}
  f\bigl(I_1,\dots,I_N\bigr) \;=\;
  \bigl(\mathbf{g}_i,\, D_i,\, P_i,\, T_i\bigr)_{i=1}^{N},
  \label{eq:vggt}
\end{equation}
where $\mathbf{g}_i$ are camera parameters, $D_i$ depth maps, $P_i \in \mathbb{R}^{H\times W\times 3}$ point maps, and $T_i$ tracking features. We use only the point-map head and its associated confidence map, discarding the other outputs. The point map assigns every pixel of view $i$ a 3D point in a common coordinate frame, yielding one point per pixel. This density of points comes with a structural limitation. VGGT predicts scene geometry in a normalized coordinate space during training, which can limit accurate representation of structures far beyond the observed camera coverage~\cite{wang2025vggt}. We observe this effect in Figure~\ref{problem_definition_image}, where the far field near the windows exhibits poor geometry, producing a blurry and hazy appearance. While this effect may be related to the training-time normalization, we do not claim to establish the exact cause.

As Figure~\ref{fig:init_coverage} shows, the two available sources of initial geometry fail in complementary regions. SfM points are measurements rather than predictions, accurate even at a distance but sparse and fewer under reduced coverage. Feed-forward dense prediction is the opposite, covering every pixel but collapsing in the far field. We therefore keep both, letting SfM anchor the far field where VGGT collapses while VGGT fills the regions SfM leaves empty. The remainder of this section describes how the two clouds are filtered, aligned, and merged.

\textbf{Confidence filtering}
VGGT predicts a per-pixel confidence alongside each point map~\cite{wang2024dust3r, wang2025vggt}. This confidence is trained jointly with the points under a confidence-weighted regression loss, so after convergence it reflects the model's expected per-pixel error. Since point confidence is not directly comparable across scenes, we apply a single threshold across all scenes. We pool the confidences from all views and retain points at or above the 50th percentile (the scene median), which by construction keeps half of the predicted points. The method is robust to this choice, as any percentile between 25 and 90 yields comparable performance, and only disabling the filter entirely degrades results (see Table~\ref{tab:ablation_conf}).

\textbf{Subsampling and Alignment}
After filtering, the dense cloud remains considerably larger than a typical SfM initialization, so we randomly subsample it to at most $400$K points, which bounds the initialization cost. The exact cap is not critical, as results are stable across a wide range (see Table~\ref{tab:ablation_cap}). The subsampled point cloud, however, still lives in a different coordinate frame. VGGT expresses its reconstruction in its own coordinate system, and SfM in another. Since vanilla 3DGS is anchored to the SfM coordinate systems through its camera poses, we align the VGGT cloud into the SfM system. Both systems process the same images, so their camera centers correspond one-to-one. We therefore estimate a similarity transform $\mathbf{x}' = sR\mathbf{x} + \mathbf{t}$ from these $N$ center pairs using the closed-form Umeyama method~\cite{umeyama1991least}, and apply it to bring the dense VGGT points into the SfM coordinate system, following standard practice in pose-free neural reconstruction and evaluation of pose estimation~\cite{lin2021barf, instantsplat, bian2023nope}.

\textbf{Union and Training}
The aligned dense cloud and the SfM point cloud are concatenated without de-duplication, reweighting, or blending. 3DGS adaptive densification and pruning naturally handle redundant points during training, so the union inherits geometry from whichever source provides it. Training proceeds from this merged representation without any other pipeline changes. Both conditions are produced with identical camera poses, architecture, optimizer, hyperparameters and held-out test poses, so the two reconstructions differ only in their initial point clouds. 

\subsection{Stereoscopic Rendering}

From the trained model, each held-out test pose is rendered in two forms. On-axis renders produces views that are directly compared with the ground truth. These are used for image metrics and the monoscopic user study. For off-axis stereo, two virtual eyes are placed with a separation corresponding to the average adult inter-pupillary distance of 6.5 cm~\cite{dodgson2004variation}. Each eye is rendered with an off-axis frustum converging on a shared image plane, the convergence distance fixed at $d_c = 1.3$ m to match the Meta Quest 2 focal distance, following the standard projection model for stereoscopic displays~\cite{southard1992transformations, Kooima2011GeneralizedPP}. Parallel camera axes are used rather than toe-in convergence to avoid introducing vertical parallax~\cite{howard2012perceiving, woods1993image}. Finally, the models are evaluated through quantitative metrics (Section~\ref{quantitative}) and a user study (Section~\ref{sec:user_study}). Observers judge the same renders under both monoscopic display and stereoscopic HMD conditions. 

\begin{figure*}[t]
\centering
\includegraphics[width=\linewidth]{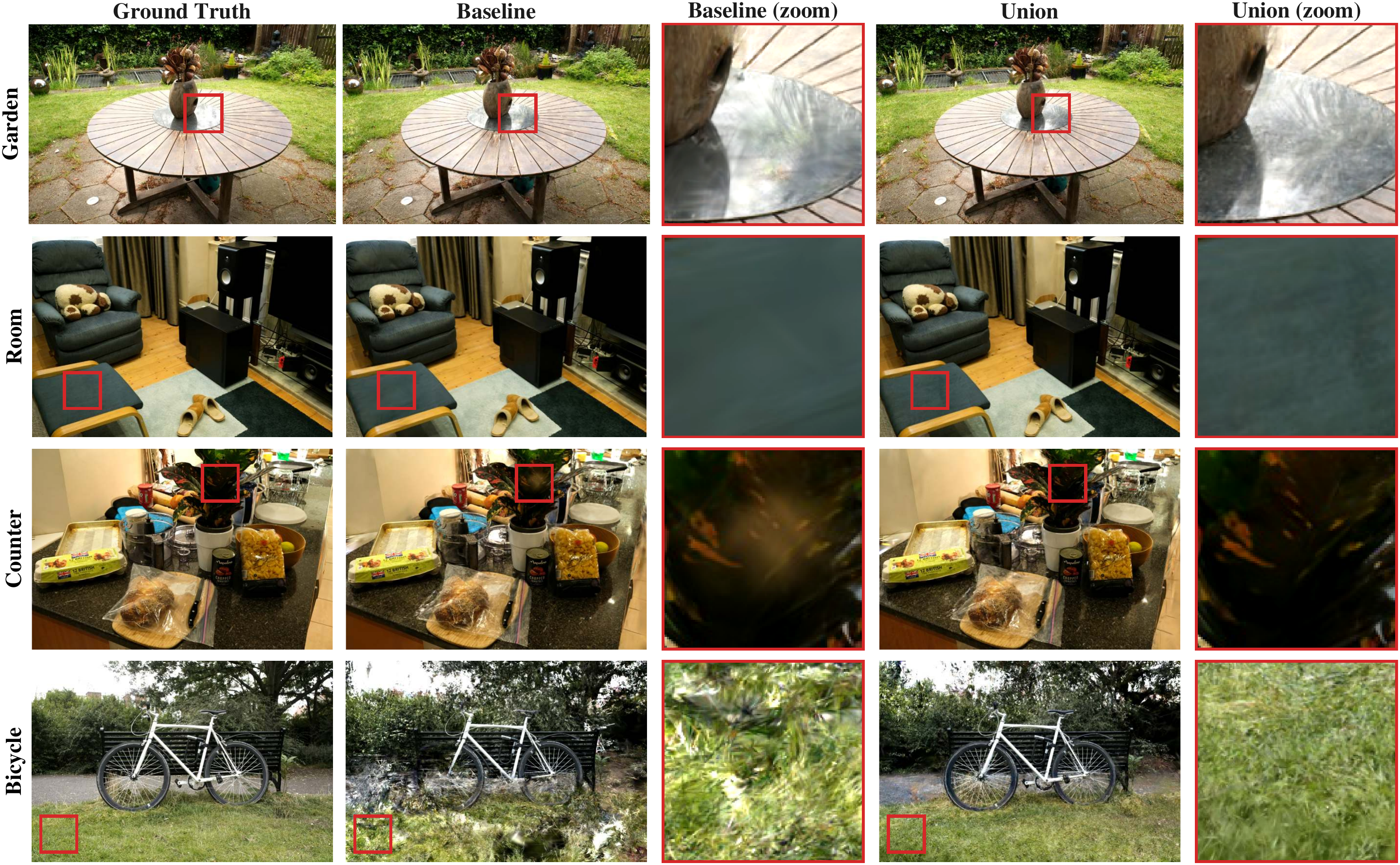}
\caption{ Ground-truth images and corresponding left-view renderings from the baseline and union initialization are shown for our user-study scenes. Red boxes indicate regions enlarged in the adjacent columns. These artifacts are examples of blurring and texture loss that is missed by the metrics and monoscopic viewing, yet becomes salient under stereoscopic viewing. }
\label{fig:metric_vs_subjective}
\end{figure*}

\subsection{Implementation Details}
We use the public 3DGS implementation~\cite{kerbl20233d} with its default optimizer, densification, and hyperparameters for both the baseline and Union initialization. All models are trained for $30{,}000$ iterations with images downsampled by a factor of four, varying by scene, on a single NVIDIA RTX 4090. Training took approximately nine minutes per scene for the baseline and a comparable time for the Union initialization. 

\subsection{Data}
We evaluate on eight scenes that are from Mip-NeRF 360 dataset~\cite{barron2022mip} and the Deep Blending dataset~\cite{DeepBlending2018}. They are spanning five indoor (room, counter, kitchen, bonsai, playroom) and three outdoor (bicycle, garden, stump). The scenes contain complex geometry and materials, including textured surfaces as well as reflective and refractive objects. 
With the dense coverage of the scene, SfM already provides adequate seed points and the reconstruction produces few salient artifacts. The benefit of our initialization is when there is reduced coverage scenarios. Hence, we uniformly subsample the input views by a fixed stride, keeping every $k$-th image for $k \in \{4, 8, 16\}$, producing training sets averaging approximately 55, 28, and 14 views respectively. A fixed stride preserves uniform angular coverage of the scene without any bias. All models are evaluated on identical held-out test views.

\section{Quantitative Evaluation}
\label{quantitative}

We evaluate the quality of the reconstruction on-axis.
For each held-out test pose, the view is rendered from the ground-truth camera position and compared against the ground-truth image using standard image metrics, namely PSNR, SSIM~\cite{wang2004image}, and LPIPS~\cite{lpips2018}, following standard practice~\cite{mildenhall2021nerf, barron2022mip, kerbl20233d}. 
We also evaluate stereoscopic quality using iSQoE~\cite{tamir2025makes}, a learned metric specifically trained on artifacts produced by 3DGS, and StereoQA~\cite{stereoqa_net}, a convolutional network that predicts perceptual stereo quality.

\subsection{Image and Stereo Quality Metrics}

Table~\ref{tab:quantitative_results} shows the quantitative results across the eight scenes for the three stride settings. Overall, the union initialization yields improvement over the SfM baseline and slight improvement over VGGT as the views become sparser. Although the magnitude varies across scenes, the gains are across PSNR, SSIM, and LPIPS.  Among the stereo-aware metrics, iSQoE shows marginal improvement between the three initialization conditions at 1/4 and 1/8 stride but StereoQA does not follow the same trend. iSQoE is specifically trained on 3DGS-rendered artifacts, while StereoQA is optimized for general stereo comfort. The artifacts lie outside the metrics training distribution, which the metrics are not trained to detect. However, these metrics alone cannot determine whether the improvement is practically meaningful. We therefore conduct a user study comparing the reconstructions under monoscopic and stereoscopic viewing conditions. We revisit these results in Section~\ref{metric_sensitivity}, where the subjective evaluations provide a reference point for assessing their significance.

\section{User Study}
\label{sec:user_study}

We conducted a user study in two viewing conditions, monoscopic viewing conditions and stereoscopic viewing conditions, following standard 2AFC protocol~\cite{series2012methodology}.
Each participant evaluated multiple scenes and each scene is evaluated by multiple participants so the observations are not independent. 
We report participant-level preference and fit a generalized linear mixed-effects model (GLMM) with a binomial logit link and crossed random intercepts for participant and scene~\cite{baayen2008mixed}.

\subsection{Setup}
Stereo pairs are rendered offline and presented in a custom WebXR application built with three.js, running on a Meta Quest 2. 
Each eye image is rendered at the native Mip-360 resolution (indoor scenes $\approx$779$\times$519 per eye, outdoor $\approx$1250$\times$830 per eye), with a fragment shader routing the left and right halves to the corresponding eye.
Session length was capped at approximately 30 minutes to limit fatigue and discomfort in the HMD. In total, we had 28 trials per participant across seven scenes at four viewpoints each. 
We excluded the stump scene because its visual artifacts were perceptually similar to those in bicycle, and we therefore retained only bicycle.
The scenes are drawn randomly from the dataset and we select the baseline and union initializations at 1/4 and 1/8 stride, where the metric difference is marginal, and investigate whether participants can still discriminate them perceptually. Example stimuli are shown in Figure~\ref{fig:metric_vs_subjective}.

\subsection{Participants.}
In total, $40$ participants took part in the experiment, $20$ each in the stereoscopic and monoscopic setting. Out of the $40$ participants, $50\%$ of them had background in visualization or computer graphics. 
They were aged $22$ to $36$, of whom $25$ were male and $15$ were female. 
We used different set of people for each of the test to avoid fatigue or carryover effects. 
Participants were rewarded for their participation.
All participants gave informed consent prior to the study. 
Authors' institution or country does not require review by an ethical review board for the type of study conducted.

\begin{figure}[t]
  \centering
  \includegraphics[width=\columnwidth]{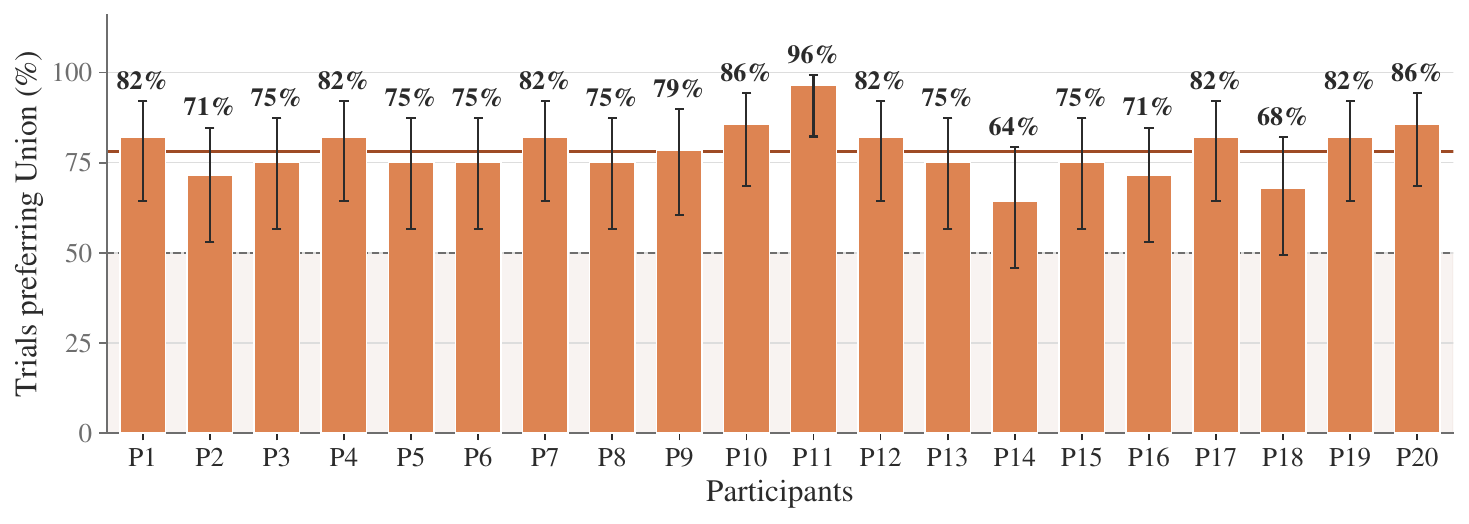}
    \caption{Per-participant preference distribution for stereoscopic viewing ($n = 20$). The x-axis shows individual participants and the y-axis shows the proportion of trials in which each participant preferred the union initialization over the baseline. Error bars are confidence intervals on that participant's 28 trials and the mean preference is $78.2\%$(95\% CI [74.8, 81.6]).}
  \label{fig:perperson_preference}

  \vspace{1.2em} 

  \includegraphics[width=\columnwidth]{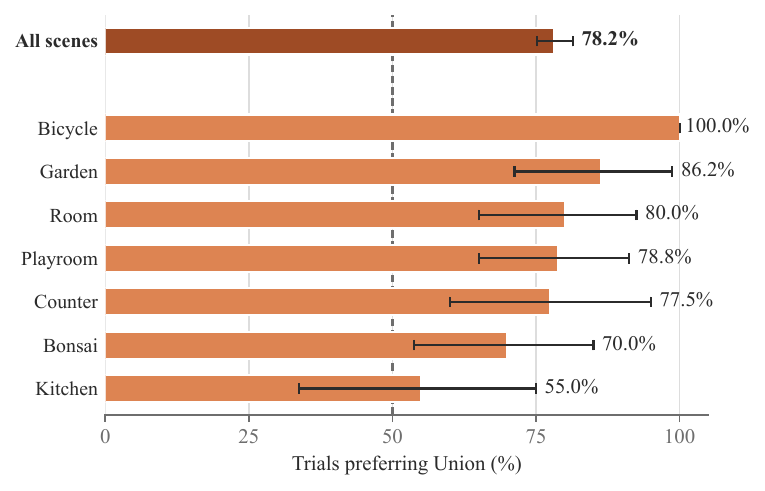}
  \caption{Per-scene preference distribution for stereoscopic viewing ($n = 20$ participants, 80 trials per scene). Bars show the percentage of trials in which the union initialization was preferred over the baseline. Error bars represent 95\% Wilson confidence intervals obtained by bootstrap participants as clusters. The mean participant preference is $78.2\%$ (95\%~CI [$75.2$, $81.4$]).}
  \label{fig:perscene_preference}
\end{figure}

\subsection{Stimuli}

\paragraph{Stereoscopic Stimili.}
The task followed a 2AFC design, consistent with standard stereoscopy visual quality assessment protocols~\cite{series2012methodology}. 
Before the main task, each participant completed a screening procedure in two parts. 
First, a random-dot stereogram to confirm that they could perceive stereoscopic depth~\cite{julesz1960binocular}. 
Second, a short tutorial with the task to demonstrate how the virtual reality interface worked. 
The full session lasted about $30$ minutes per participant and included three breaks of one minute each to reduce fatigue. 
In each trial the participant viewed a pair of stimuli of the same scene and viewpoint, one rendered from the baseline and Union initialization, with the two methods anonymized. 
Trial order was randomized per participant, with the two method presentation counterbalanced (half saw baseline first, half saw Union first). 
Scene order was independently randomized for each participant. 
A uniform grey screen was displayed for $0.5$ seconds between trials to reset visual adaptation, with a longer reset of $1.5$ applied between scenes. 
Participants were asked: ``Which stereo pair looks more realistic? ''. There was no time restriction for the participants.

\paragraph{Monoscopic Stimuli.}
For the monoscopic condition, a separate group of 20 participants (not overlapping with the stereoscopic sample) judged the same reconstructions using 2AFC assessment protocol. 
Each scene was rendered on-axis from a single viewpoint per trial at the same per-eye resolution used in the stereoscopic condition. 
Indoor scenes were $779\times519$ and outdoor scenes were approximately $1250\times830$. 
The renders were presented as static non-stereoscopic image pairs on a 27-inch $2560\times1440$ display at approximately $60$\,cm. 
Stimuli were drawn from the same seven scenes and camera views as the stereoscopic condition. 
Only the participant sample differed between conditions. 
The underlying renders and viewpoints were identical. 
Trial order, side of the scene, and baseline versus Union presentation were randomized and counterbalanced per participant.  
Participants were asked ``Which image looks more realistic?''. 
Participants did not have any time restriction to choose between the pairs.

\begin{figure}[t]
  \centering
  \includegraphics[width=\columnwidth]{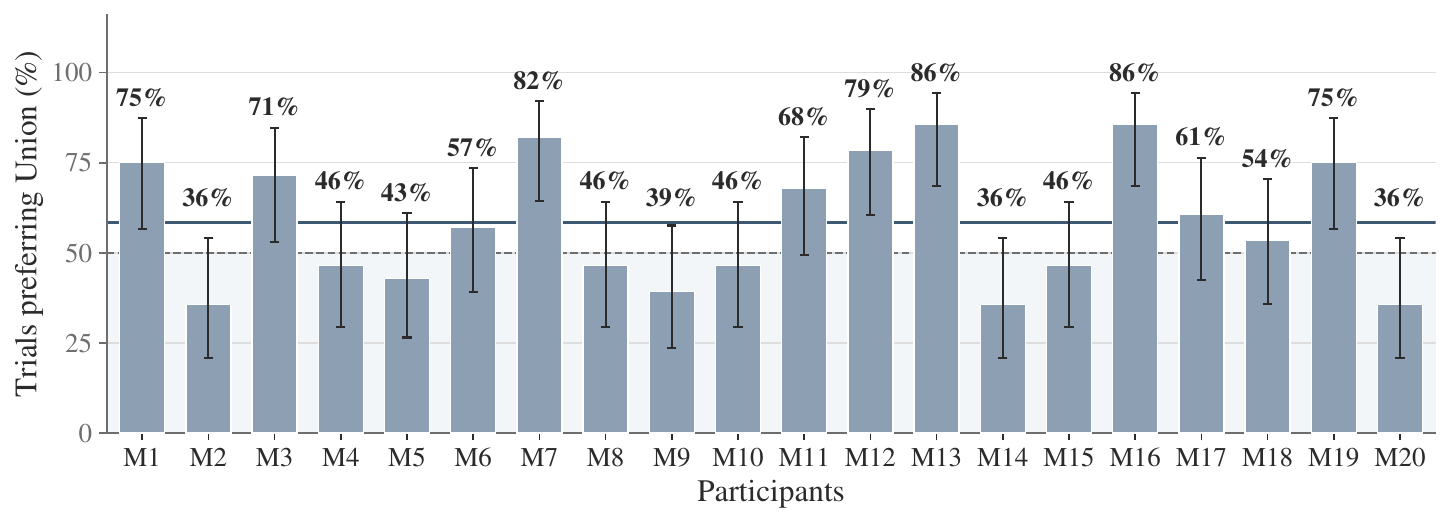}
    \caption{Per-participant preference distribution for monoscopic viewing ($n = 20$). The x-axis shows individual participants and the y-axis shows the proportion of trials in which each participant preferred the Union initialization over the baseline. Error bars are 95\% confidence intervals on that participant's 28 trials and the mean preference is $58.4\%$ (95\% CI [50.1, 66.7])}
  \label{fig:perperson_preference_mono}

      \vspace{1.2em} 

        \includegraphics[width=\columnwidth]{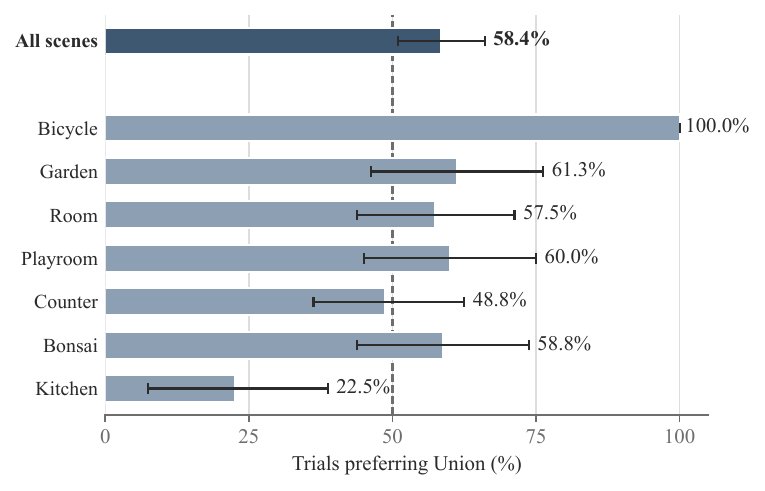}
    \caption{Per-scene preference distribution for monoscopic viewing ($n = 20$ participants, 80 trials per scene). Bars show the percentage of trials in which the union initialization was preferred over the baseline. Error bars represent 95\% confidence intervals obtained by bootstrap participants as clusters. The mean participant preference is $58.4\%$ (95\%~CI [$50.9$, $66.1$]).}
  \label{fig:perscene_preference_mono}
\end{figure}

\subsection{Results}

\paragraph{Stereoscopic Results}
We collected 560 responses in the stereoscopic condition (20 participants, 28 trials each). 
The union initialization was preferred in 438 of 560 trials (78.2\%).  
The mean participant preference was 78.2\% (95\% CI [74.8, 81.6], $n = 20$), as shown in Figure~\ref{fig:perperson_preference}.
The GLMM gives a conditional preference of 82.0\% (95\% CI [78.6, 84.9]), significantly above chance (OR = 4.54, 95\% CI [3.68, 5.61], $p < 0.001$).
Per-scene preference is shown in Figure~\ref{fig:perscene_preference} and computed 95\% confidence interval by resampling participant as clusters.
The union initialization was preferred in every scene, with six of the seven scenes significantly above chance. Only \textit{kitchen} (55.0\% ) scene did not differ from chance.


\paragraph{Monoscopic Results}
We collected 560 responses in the monoscopic condition (20 participants, distinct from the stereoscopic group, 28 trials each).
The union initialization was preferred in 327 of 560 trials (58.4\%).
The mean participant preference was 58.4\% (95\% CI [50.1, 66.7], $n = 20$), as shown in Figure~\ref{fig:perperson_preference_mono}.
The GLMM gives a conditional preference of 66.0\% (95\% CI [61.4, 70.3]), significantly above chance (OR = 1.94, 95\% CI [1.59, 2.36], $p < 0.001$).
For the per-scene analysis, we used the same 95\% confidence interval procedure as before. Only \textit{bicycle} was preferred above chance for the union, and only \textit{kitchen} was preferred above chance for the baseline. The remaining scenes did not differ from chance (Figure~\ref{fig:perscene_preference_mono}).
This indicates that the monoscopic viewing the two reconstructions can be distinguished accurately only where the artifact is most severe and are otherwise not dependably separated.

\paragraph{Participant Agreement}

The two conditions differ not only in effect size but in the distribution of variance across observers and scenes.
In the stereoscopic condition, between-scene variance ($\sigma_{\text{scene}} = 1.18$, 95\%~CI [0.72, 1.95]) substantially exceeds between-participant variance ($\sigma_{\text{participant}} = 0.27$ [0.19, 0.37]). 
This indicates that participants agree closely with one another, and the effect size is determined primarily by scene content.
In the monoscopic condition, between-scene variance ($\sigma_{\text{scene}} = 1.83$ [1.12, 2.98]) remains larger than between-participant variance ($\sigma_{\text{participant}} = 0.91$ [0.67, 1.23]), but the gap narrows considerably.  
The scene-to-participant variance ratio falls from $\sim$4.4$\times$ stereoscopically to $\sim$2.0$\times$ monoscopically, with between-participant variance more than three times the stereoscopic value.
This pattern is consistent with clearly visible artifacts in the stereoscopic condition and marginal artifacts in the monoscopic condition.
When a difference is clearly visible, observers converge on the same judgment and residual variance reflects stimulus properties.
When it is marginal, observers fall back on individual criteria and their responses scatter. We note, however, that $\sigma_{\text{scene}}$ is estimated from only seven scenes and its interval is correspondingly wide.

\paragraph{Metric Sensitivity}
\label{metric_sensitivity}

On the 28 frames used in the study, the two initializations differ by $+0.77$~dB PSNR and $+0.030$ SSIM, while LPIPS is reduced by $0.040$.
The stereo-aware metrics differ by $-0.015$ for iSQoE and $-0.69$ for StereoQA.
Across five training runs of the baseline differing only in random seed, all the metrics showed marginal variation within fixed configuration (supplemental Table~\ref{tab:seed_variance}).
Despite these small metric differences, observers preferred the union initialization in $78.2\%$ of stereoscopic trials, and every participant preferred it in a majority of trials.
This contrast indicates that existing metrics under-represent localized artifacts that become salient under stereoscopic viewing.

\begin{table}[t]
\centering
\caption{The 28 frames used in the user study (four per scene), scored with standard metrics. Stereo-aware metrics showed marginal and inconsistent differences, so we report iSQoE as the representative stereo metric. \textbf{Bold} indicates better objective performance or a subjective preference.}
\label{tab:study_frames}
\resizebox{\columnwidth}{!}{%
\setlength{\tabcolsep}{3pt}
\begin{tabular}{l cccccccc cc}
\toprule
\multirow{2}{*}{Scene} & \multicolumn{2}{c}{PSNR$\uparrow$} & \multicolumn{2}{c}{SSIM$\uparrow$} & \multicolumn{2}{c}{LPIPS$\downarrow$} & \multicolumn{2}{c}{iSQoE$\downarrow$} & \multicolumn{2}{c}{Pref. (\%)$\uparrow$} \\
\cmidrule(lr){2-3}\cmidrule(lr){4-5}\cmidrule(lr){6-7}\cmidrule(lr){8-9}\cmidrule(l){10-11}
 & SfM & Union & SfM & Union & SfM & Union & SfM & Union & Mono & VR \\
\midrule
bicycle 1/8  & 14.73 & \textbf{18.23} & 0.226 & \textbf{0.398} & 0.485 & \textbf{0.327} & 0.662 & \textbf{0.591} & \textbf{100.0} & \textbf{100.0} \\
bonsai 1/4   & \textbf{26.49} & 26.35 & \textbf{0.910} & 0.909 & 0.076 & \textbf{0.062} & 0.482 & \textbf{0.481} & 58.8 & \textbf{70.0} \\
counter 1/8  & 21.92 & \textbf{22.23} & 0.764 & \textbf{0.773} & 0.169 & \textbf{0.129} & \textbf{0.576} & 0.580 & 48.8 & \textbf{77.5} \\
garden 1/8   & 19.94 & \textbf{20.69} & 0.578 & \textbf{0.618} & 0.224 & \textbf{0.176} & 0.492 & \textbf{0.456} & 61.3 & \textbf{86.2} \\
kitchen 1/8  & \textbf{23.80} & 22.56 & \textbf{0.855} & 0.808 & \textbf{0.094} & 0.137 & \textbf{0.524} & 0.572 & 22.5 & \textbf{55.0} \\
playroom 1/4 & 23.69 & \textbf{24.97} & 0.857 & \textbf{0.883} & 0.201 & \textbf{0.155} & 0.745 & \textbf{0.702} & 60.0 & \textbf{78.8} \\
room 1/4     & 29.52 & \textbf{30.45} & 0.928 & \textbf{0.940} & 0.079 & \textbf{0.060} & 0.503 & \textbf{0.495} & 57.5 & \textbf{80.0} \\
\midrule
\textbf{Mean} & 22.87 & \textbf{23.64} & 0.731 & \textbf{0.761} & 0.190 & \textbf{0.149} & 0.569 & \textbf{0.554} & 58.4 & \textbf{78.2} \\
\bottomrule
\end{tabular}%
}
\end{table}

\section{Discussion}

Standard metrics evaluate the reconstruction by averaging across pixels and views, but this process smooths over localized errors and misses penalizing them. 
Stereo-aware metrics trained on 3DGS artifacts also fail to distinguish them.
The same geometric error is perceived differently in the two viewing conditions.
On a monoscopic screen, a misplaced floater produces no disparity error at all.
It appears as blur or texture duplication, which observers ignore when judging image quality. 
In stereo, the same geometry is rendered at the wrong disparity in both eyes and as seen as a misplaced blob.
Changes between the conditions are depth placement, not image fidelity, and the visual system is sensitive to errors of this kind~\cite{kooi2004visual, zhu2025perceptual}. 
A practitioner selecting a reconstruction for VR on these measures would find no basis to prefer one initialization over the other, yet observers in the headset do so consistently.

Two further observations emerged from the user study.  
First, legibility acted as a strong cue. A reconstruction rendering text or fine markings readably was preferred even in the presence of other minor artifacts. 
Second, participants appeared to weight near-field content most heavily, which amplifies near-field errors.
This is because a geometry error at close range produces a far larger disparity error than the same error at distance.

The one non-significant scene is the indoor \textit{kitchen} (55.0\%). 
The VGGT network places a black streak of near-field geometry in front of the scene, and because the union initialization retains all points, it inherits this floater as well. 
The union initialization helps significantly with outdoor scenes where the deficit is missing far-field coverage, as in \textit{bicycle} (100\%), and with indoor scenes where SfM leaves gaps, as in \textit{room} (80\%) and \textit{bonsai} (70\%). The union therefore helps when geometry is missing from the initialization but cannot correct artifacts already present in it.

\section{Ablation Study}
\label{sec:ablation_study}

We run a parameter ablation study to validate the parameters chosen in the union initialization.
We evaluate the confidence threshold $\tau$ used to filter VGGT points at different percentiles, and the initialization size with different point caps. 
We interpret this result as a sensitivity analysis rather than evidence of statistical significance.

\subsection{Confidence-threshold ablation}

VGGT assigns a confidence score to each predicted point, but not all points are geometrically reliable. 
We therefore discard points below a percentile threshold before computing the union, removing unreliable geometry while preserving dense coverage. We set $\tau = p_{50}$ a priori as a midpoint and retain the higher-confidence half of the points.
Table~\ref{tab:ablation_conf} reports performance from no filtering ($p_0$) up to an aggressive $p_{90}$. 
Across this range, PSNR varies by 1.2 dB on \textit{room} and 0.71 dB on \textit{kitchen}, SSIM by 0.11 and 0.021, and LPIPS by 0.007 and 0.026, respectively. 
No single confidence threshold dominates across scenes and metrics. 
This ablation therefore does not identify an optimal threshold, but indicates that the choice of threshold is scene- and metric-dependent.

\begin{table}
  \centering
  \caption{Confidence-threshold ablation (cap $=400$k). VGGT point are kept above confidence percentile $p$ ($0$ to $90$). Quantitative scores remain stable across thresholds.  }
  \label{tab:ablation_conf}
  \scriptsize
  \setlength{\tabcolsep}{3pt}
  \begin{tabular}{l ccc ccc}
  \toprule
  & \multicolumn{3}{c}{room 1/4} & \multicolumn{3}{c}{kitchen 1/8} \\
  \cmidrule(lr){2-4}\cmidrule(lr){5-7}
  $p$ & PSNR & SSIM & LPIPS & PSNR & SSIM & LPIPS \\
  \midrule
  SfM        & 29.82 & 0.924 & 0.085 & 23.08 & 0.847 & 0.128 \\
  $0$        & 30.85 & 0.930 & 0.073 & 23.95 & 0.854 & 0.109 \\
  $25$       & 30.66 & 0.928 & 0.074 & 23.82 & 0.833 & 0.133 \\
  $50$        & 30.52 & 0.925 & 0.074 & 23.66 & 0.834 & 0.125 \\
  $75$       & 30.10 & 0.922 & 0.077 & 24.14 & 0.849 & 0.111 \\
  $90$       & 29.60 & 0.919 & 0.080 & 24.27 & 0.851 & 0.113 \\
  \bottomrule
  \end{tabular}
\end{table}

\subsection{Initialization-size}

The maximum number of VGGT points clouds is the second hyperparameter in our Union initialization.
We set this cap to 400k points a prio to evaluate the effect. 
The Initialization sizes range from $25$k to $1.6$M, no single cap consistently performs best across scenes and all metrics. 
On \textit{room}, PSNR varies by $0.42$~dB, SSIM by $0.11$, and LPIPS by $0.007$. 
On \textit{kitchen}, PSNR varies by $0.51$~dB, SSIM by $0.017$, and LPIPS by $0.023$.
Performance does not vary monotonically with initialization size. 
These results indicate limited, scene-dependent sensitivity to the point cap within the tested range. 
We therefore retain $400$k as the default that provides dense coverage without the added computation of a larger initialization.

  \begin{table}
  \centering
  \caption{Initialization seed point ablation study. Quantitative scores remain stable across a wide range of initialization sizes.}
  \label{tab:ablation_cap}
  \scriptsize
  \setlength{\tabcolsep}{3pt}
  \begin{tabular}{l ccc ccc}
  \toprule
  & \multicolumn{3}{c}{room 1/4} & \multicolumn{3}{c}{kitchen 1/8} \\
  \cmidrule(lr){2-4}\cmidrule(lr){5-7}
  cap & PSNR & SSIM & LPIPS & PSNR & SSIM & LPIPS \\
  \midrule
  SfM         & 29.82 & 0.924 & 0.085 & 23.08 & 0.847 & 0.128 \\
  25k         & 30.32 & 0.928 & 0.081 & 23.91 & 0.846 & 0.123 \\
  50k         & 30.48 & 0.929 & 0.079 & 23.44 & 0.836 & 0.137 \\
  100k        & 30.36 & 0.930 & 0.076 & 23.60 & 0.843 & 0.129 \\
  200k        & 30.10 & 0.927 & 0.076 & 23.81 & 0.838 & 0.128 \\
  400k        & 30.52 & 0.925 & 0.074 & 23.66 & 0.834 & 0.125 \\
  800k        & 30.23 & 0.923 & 0.076 & 23.73 & 0.840 & 0.114 \\
  1600k       & 30.29 & 0.919 & 0.077 & 23.48 & 0.829 & 0.133 \\
  \bottomrule
  \end{tabular}
  \end{table}


\section{Limitations and Future Work}

Our study and evaluation have several limitations. The study evaluated static stereo image pairs displayed in an HMD rather than free six-degree-of-freedom (6-DoF) exploration with head motion. Consequently, we characterized how artifacts in binocular viewing affect perceived quality, but not how they behave under continuous viewpoint change. Future work includes extending the evaluation to free-viewpoint 6-DoF navigation, measuring comfort and presence during head motion, and investigating whether temporal artifacts or depth inconsistencies emerge under interactive exploration.

The two viewing conditions were run with two separate groups of participants on different hardware, so the comparison between them reflects the viewing condition as a whole rather than the stereoscopic condition in isolation. We chose separate groups because a within-subject design would go beyond a comfortable duration for a study. 

The user study included only pairwise comparisons between the baseline and hybrid initialization. Adding VGGT would have tripled the pairwise comparisons per scene and extended each session beyond a comfortable duration in an HMD. The extension is a three-way perceptual comparison that also includes VGGT, split across shorter sessions, which would reveal the differences between baseline, VGGT, and Union in stereo.

Our initialization strategy combines a union of SfM and the feed-forward network VGGT. The limitations of both sources propagate directly into the fused representation, and artifacts inherent to either SfM or VGGT may persist in the final reconstruction. A promising direction for future work is to develop an adaptive fusion mechanism that weights contributions by scene-space proximity or per-point confidence, suppressing inherited artifacts without sacrificing the complementary coverage gained by combining both sources.

\section{Conclusion}

Our work shows that stereoscopic viewing exposes 3DGS artifacts which monoscopic viewing and standard image metrics fail to resolve. 
We constructed stimulus pairs differing only in initialization, so the difference observers respond to is the reconstructed geometry.
The difference between initializations is marginal within the metrics and under monoscopic viewing, but under stereoscopic viewing the same pair of reconstructions is consistently preferred.
Current metrics are not designed to capture localized errors of this kind and are difficult to perceive under monoscopic viewing. 
If the intended use case is VR, reconstruction should be evaluated perceptually, in stereo, in the headset.

\acknowledgments{}

\bibliographystyle{abbrv-doi}
\bibliography{template}

\section*{Supplemental Materials}
\label{sec:supplemental_materials}

\begin{table*}[!t]
\centering
\caption{Across five baseline training runs differing only in random seed. The mean run-to-run standard deviations are PSNR~$0.37$~dB, SSIM~$0.010$, LPIPS~$0.011$, iSQoE~$0.013$, and StereoQA~$0.28$ (maxima: PSNR~$0.72$~dB, StereoQA~$0.39$).}
\label{tab:seed_variance}
\small
\setlength{\tabcolsep}{4pt}
\begin{tabular}{lccccc}
\toprule
Scene / split & PSNR ($\pm$std) & SSIM ($\pm$std) & LPIPS ($\pm$std) & iSQoE ($\pm$std) & StereoQA ($\pm$std) \\
\midrule
room 1/4      & 29.65 (0.18) & 0.922 (0.001) & 0.086 (0.001) & 0.559 (0.002) & 22.03 (0.24) \\
room 1/8      & 23.59 (0.44) & 0.799 (0.009) & 0.197 (0.008) & 0.683 (0.008) & 22.48 (0.25) \\
bicycle 1/4   & 20.60 (0.15) & 0.460 (0.006) & 0.404 (0.007) & 0.564 (0.009) & 13.54 (0.34) \\
bicycle 1/8   & 15.68 (0.72) & 0.234 (0.035) & 0.486 (0.031) & 0.656 (0.050) &  8.34 (0.22) \\
kitchen 1/4   & 27.71 (0.46) & 0.916 (0.006) & 0.070 (0.008) & 0.453 (0.004) & 17.80 (0.13) \\
kitchen 1/8   & 24.03 (0.26) & 0.850 (0.005) & 0.121 (0.008) & 0.500 (0.006) & 18.58 (0.39) \\
\midrule
\textbf{Mean} & \textbf{0.37} & \textbf{0.010} & \textbf{0.011} & \textbf{0.013} & \textbf{0.28} \\
\textbf{Max}  & \textbf{0.72} & \textbf{0.035} & \textbf{0.031} & \textbf{0.050} & \textbf{0.39} \\
\bottomrule
\end{tabular}
\end{table*}

\end{document}